\documentclass[aps,pra,twocolumn,showpacs,superscriptaddress,floatfix]{revtex4}
\usepackage{graphicx}
\usepackage{amsmath}
\begin{document}

\title{Quantum Dissipative Dynamics of Entanglement in the Spin-Boson Model}
\author{Jong Soo Lim}
\affiliation{Department of Physics, Seoul National University, Seoul
  151-747, Korea}
\author{Mahn-Soo Choi}
\email{choims@korea.ac.kr}
\affiliation{Department of Physics, Korea University, Seoul 136-701, Korea}
\affiliation{Department of Physics and Astronomy, University of Basel,
  4056 Basel, Switzerland}
\author{M. Y. Choi}
\affiliation{Department of Physics, Seoul National University, Seoul
  151-747, Korea}
\affiliation{Korea Institute for Advanced Study, Seoul 130-722, Korea}

\begin{abstract}
We study quantum dissipative dynamics of entanglement in the spin-boson
model, described by the generalized master equation.  We consider the
two opposite limits of pure-dephasing and relaxation models, measuring
the degree of entanglement with the concurrence.  When the Markovian
approximation is employed, entanglement is shown to decay exponentially
in both cases.  On the other hand, non-Markovian contributions alter the
analytic structure of the master equation, resulting in logarithmic
decay in the pure dephasing model.
\end{abstract}
\maketitle

\let\up=\uparrow
\let\down=\downarrow
\newcommand\varD{\mathcal{D}}
\newcommand\varL{\mathcal{L}}
\newcommand\varP{\mathcal{P}}
\newcommand\varQ{\mathcal{Q}}
\newcommand\half{\frac{1}{2}}
\newcommand\tr{\mathrm{Tr}}
\newcommand\ket[1]{\left|{\textstyle #1}\right\rangle}
\newcommand\bra[1]{\left\langle{\textstyle #1}\right|}

The dynamical properties of an open quantum system are conveniently
characterized by its decoherence times, viz, how long the system can
maintain coherence in a given superposition state.  It is customary
and useful to distinguish further two limiting decoherence processes, relaxation and dephasing, 
although realistic decoherence phenomena appear with both processes admixed.
In many cases of quantum information processing, a quantum device consists of two-level systems, 
for which rates of the two processes are designated by $1/T_1$ and $1/T_2$, respectively, 
in the convention of nuclear magnetic resonance. 
When the system can be partitioned into two parts, one can exploit
another interesting property of the system, the entanglement between the
two parts.
As highlighted by the EPR paradox \cite{Einstein35a} and Bell's theorem \cite{Bell66a,Greenberger90a}, entanglement has been regarded as one of the most fundamental properties of quantum physics.  
In recent decades, entanglement has attracted renewed interest as a key resource
for quantum information processing such as quantum dense coding \cite{Bennett92a}, 
quantum teleportation \cite{Bennett93a}, and quantum computation \cite{Shor94a}.
Since the first experimental generation of the entangled photon pairs \cite{Aspect82a}, 
a great number of researches have been devoted to the reliable creation and maintenance
of the entanglement, e.g., between far-separated photons \cite{Tittel98a}, 
between solid-state qubits \cite{Yamamoto03a,ChoiMS00c},
and beween macroscopic atomic ensembles \cite{Julsgaard01a}.

A natural question is then how the entanglement between the two
partitions evolves in time.
In particular, it will be useful to examine if
such dynamics of the entanglement can be characterized by a separate
time scale and if so, how it is related to the decoherence times (e.g.,
$T_1$ and $T_2$) of the system.
In this paper, we investigate the dissipative dynamics of two ``spins''
(two-level systems) coupled to environment within the spin-Boson model
both in the Markovian and non-Markovian limits, and examine the
characteristic time scale over which the entanglement decays.  It is
found that entanglement may decay exponentially or logarithmically,
depending on the presence of spin relaxation.

%
We consider two spins, 
coupled respectively to separate (``local'') baths of harmonic
oscillators\cite{endnote1}.
Following Ref. \onlinecite{Leggett87a}, we adopt the spin-boson Hamiltonian
\begin{equation}
H  = H_S + H_B + H_{SB},
\end{equation}
with
\begin{equation}
\label{entangle::eq:1}
H_S = \half \sum_{j=1,2} \left(\epsilon_j\sigma_j^z + \Delta_j\sigma_j^x\right)
\end{equation}
describing isolated spins,
\begin{equation}
\label{entangle::eq:5}
H_B = \sum_{j=1,2}\sum_{m} \omega_{j,m}b_{j,m}^\dag b_{j,m}^{}
\end{equation}
baths of harmonic oscillators, and
\begin{equation}
\label{entangle::eq:6}
H_{SB} = \sum_{j=1,2} \sum_{m} (g_{j,m}\sigma_j^z b_{j,m}^\dag +
g_{j,m}^*\sigma_j^{z\dag} b_{j,m}^{})
\end{equation}
coupling between the spins and the oscillators, where $\sigma_j^\mu$
($j=1,2$ and $\mu=x,y,z$) are Pauli matrices representing the $j$th spin
and $b_{j,m}^\dag$/$b_{j,m}^{}$ are creation/annihilation operators for
the oscillator of mode $m$ coupled to the $j$th spin.  The unit system
$\hbar=k_B=1$ is used throughout this work.  The dynamics of the spins
at time scales longer than the bath correlation time $\omega_c^{-1}$,
which is of our concern, does not depend on the details of the coupling
constants $g_{j,m}$ and the oscillator frequencies $\omega_{j,m}$.  It
suffices to characterize each bath collectively in terms of the bath
spectral density function \cite{Leggett87a}
\begin{equation}
\label{entangle::eq:3}
J_j(\omega) \equiv \sum_m |g_{j,m}|^2 \delta(\omega-\omega_{j,m}).
\end{equation}
In this work, we focus on the Ohmic model of dissipation
\begin{equation}
\label{ohmic}
J_j(\omega) = \half\alpha_j\omega e^{-\omega/\omega_c} ,
\end{equation}
where $\alpha_j$ is the dimensionless damping parameter of the bath
coupled to the $j$th spin. 

In spite of the simple form of the Hamiltonian, the dynamics of the
spin-Boson model is highly non-trivial and has been the subject of a
number of works \cite{Leggett87a}.  Note in particular that the two
spins in the system are completely decoupled with each other, each
interacting separately with its \emph{local} bath.
In equilibrium, the density matrix $\rho_S$ of the two spins is
therefore separable:
$\rho_S(t\to\infty)=\rho_1(\infty)\otimes\rho_2(\infty)$.  However,
starting from an initial state $\rho_S (0)$ with a finite amount of
entanglement, the system displays non-trivial time evolution of the
entanglement contents in $\rho_S (t)$.  [See
Refs. \onlinecite{Aspect82a,Tittel98a,Yamamoto03a,ChoiMS00c} for initial
preparation of entanglement between two spins.]

To proceed, we diagonalize Eq.~\eqref{entangle::eq:1}, and obtain 
\begin{eqnarray}
H_S &=& \half \sum_{j=1,2} E_j \tau_j^z \\
H_{SB} &=& \sum_{j=1,2}\sum_{m} 
\left[ \tau_j^z\cos\theta_j\left(g_{j,m}b_{j,m}^{\dag} +
    g_{j,m}^{\ast}b_{j,m}\right)\right. \nonumber \\
 & & ~~~~~\left. + \tau_j^x\sin\theta_j\left(g_{j,m}b_{j,m}^{\dag} +
     g_{j,m}^{\ast}b_{j,m}\right) \right],
\end{eqnarray}
where $E_j = \sqrt{\epsilon_j^2 + \Delta_j^2}$, $\theta_j =
\tan^{-1}(\Delta_j/\epsilon_j)$, and $\tau_j^\mu$ ($\mu=x,y,z$) are the
Pauli matrices in the rotated basis.  Here we probe the evolution of
entanglement in the two limiting cases: When
$|\epsilon_j|\gg|\Delta_j|$, i.e, $\theta_j \approx 0$,
each spin is not allowed to flip its direction: $[\tau_j^z,H]=0$.  In
the analogy to a particle in a double-well potential \cite{Leggett87a},
this corresponds to the case of an infinitely high potential barrier.
The presence of the bath thus cannot change the population of spin
states, merely breaking the coherent superposition of different spin
states.  In this sense we call this limit the ``pure dephasing'' model
\cite{Preskill98a}.  In the opposite limit $|\epsilon_j|\ll|\Delta_j|$,
i.e, $\theta_j \approx \pi/2$,
with the rotating wave approximation employed, coupling to the bath
leads to relaxation of spin polarizations; this corresponds to the
strong bias potential in the analogy to the particle in a double-well
potential.  We thus call this limit the ``relaxation" model.

The evolution of the density matrix $\rho$ of the total (isolated) system
is unitary and follows the von Neumann equation
\begin{eqnarray}
\label{entangle::eq:2}
\frac{d\rho}{dt} = -i[H,\rho] \equiv \varL\rho ,
\end{eqnarray}
with the Liouville operator $\varL = \varL_S +\varL_B +\varL_{SB}$
corresponding to the three parts of $H$.  The reduced density matrix
$\rho_S=\tr_{B}\rho$ of spins, given by the trace of $\rho$ over the
bath variables $B$, is not unitary due to the coupling to baths.  Its
equation of motion, obtained from Eq.~(\ref{entangle::eq:2}) by taking
the trace, reads
\begin{equation}
\label{entangle::eq:4}
\frac{d\rho_S}{dt} = \varL_S\rho_S + \varD\rho_S 
\end{equation}
with 
the dissipative part formally expressed as \cite{Zwanzig60a}
\begin{equation}
\label{master}
\varD\rho_S(t) =
- \int_0^t dt' \,\tr_B
\left[H_{SB}, e^{\varQ\varL(t-t')} 
       \left[H_{SB},\rho_S(t')\otimes\rho_B\right]
\right] ,
\end{equation}
where we have introduced the projection operators
$\varP\rho\equiv(\tr_B\rho)\otimes\rho_B$ and $\varQ \equiv 1-\varP$
with the equilibrium bath density matrix $\rho_B \equiv
e^{-H_B/T}/\tr_Be^{-H_B/T}$ at temperature $T$ \cite{Zwanzig60a}.
We have also assumed $\rho(0)=\rho_S(0)\otimes\rho_B$
for the initial configuration .
Further, to get analytic expressions for $\rho_S(t)$, we take the
first-order Born approximation and make the replacement in
Eq.~(\ref{master})
\begin{equation}
\label{born}
e^{\varQ\varL(t-t')} \to e^{\varQ(\varL_S + \varL_B)(t-t')} ,
\end{equation}
which is valid for sufficiently weak spin-bath coupling.

We measure the entanglement degree in the mixed state $\rho_S$ by means of
the \emph{concurrence}
\begin{equation}
\label{conc}
C(\rho_S)
\equiv \max \left[ \sqrt{\lambda_1} - \sqrt{\lambda_2}
  - \sqrt{\lambda_3} - \sqrt{\lambda_4},\, 0 \right],
\end{equation}
where $\lambda_1 \ge \lambda_2 \ge \lambda_3 \ge \lambda_4$ are the
eigenvalues of the matrix $\rho_S (\tau_{y} \otimes \tau_{y})
\rho_S^{\ast} (\tau_y \otimes \tau_y)$.  The concurrence vanishes
for a separable state and becomes unity for a maximally entangled state.
While it is possible to investigate $C(\rho_S(t))$ for a general initial
state $\rho_S(0)$, we shall usually assume, for clearer physical
interpretation, a particular class of initial states, namely the Werner
states:
\begin{equation}
\label{werner}
\rho_S(0) = W(r) \equiv r|\Phi^{+}\rangle\langle\Phi^{+}| + \frac{1-r}{4}\,I ,
\end{equation}
where
\begin{math}
\ket{\Phi^+}
=\frac{1}{\sqrt{2}} \left(\ket\up\otimes\ket\down
  +\ket\down\otimes\ket\up\right)
\end{math}
is one of the Bell states and $I$ denotes the $4\times 4$
identity matrix.  The parameter $r$ in the range $[0, 1]$ interpolates
between the fully random state $I/4$ (for
$r=0$) and the maximally entangled state $\ket{\Phi^+}\bra{\Phi^+}$ (for $r=1$).

%
The master equation for $\rho_S$, given by
Eqs.~(\ref{entangle::eq:4})--(\ref{born}), takes the form of an
integro-differential equation, and the state $\rho_S(t)$ at time $t$
depends on the history $\rho_S(t')$ with $t'\leq t$.  We first discuss
the Markovian limit,
where bath correlations decay sufficiently fast compared with the
relaxation time of the spins.  Mathematically, it corresponds to
replacing $\rho_S(t')$ in the integrand of Eq.~(\ref{master}) by
$\rho_S(t)$.  With this Markovian approximation, the master
equation~(\ref{entangle::eq:4}) reduces to a linear differential
equation with time-independent coefficients.  Then expected is
exponential behavior of $\rho_S(t)$ and accordingly of the concurrence
$C[\rho_S(t)]$, from which one can extract the characteristic time scale
of the entanglement dynamics.

For the pure dephasing model ($\Delta_j=0$), the dissipative part given
by Eq.~(\ref{master}) reads
\begin{equation}
\label{entangle::eq:7}
\varD\rho_S(t) =
\sum_j\gamma_j\left[\tau_j^z\rho_S(t)\tau_j^z-\rho_S(t)\right],
\end{equation}
where
\begin{math}
\gamma_j=2\pi\alpha_j T
\end{math}
is the dephasing rate of the $j$th spin.
Given an initial state $\rho_S(0)=W(r)$, it is straightforward to find
the solution $\rho_S(t)$ of Eqs.~(\ref{entangle::eq:4}) and
(\ref{entangle::eq:7}).  It leads to the concurrence
\begin{equation}
\label{entangle::eq:8}
C(r,t) = \Theta\left(r-\frac{1}{3}\right)
\left[r e^{-4(\gamma_1+\gamma_2)t} - \frac{1-r}{2}\right],
\end{equation}
where $\Theta(x)$ is the unit step function.
In particular, for the
maximally entangled initial state ($r=1$), the concurrence becomes
\begin{eqnarray}
C(r{=}1,t) &=& e^{-4(\gamma_1+\gamma_2)t} ,
\end{eqnarray}
which decays exponentially with the rate proportional to the dephasing
rate $\gamma_1+\gamma_2$.  Notice that the concurrence does not decay at
zero temperature, where the dephasing rates vanishes.
%

For the relaxation model, Eq.~(\ref{master}) reduces to
\begin{widetext}
\begin{equation}
\label{entangle::eq:9}
\varD\rho_S(t) = \sum_{j=1,2}\Bigg\{
  i\delta E_j\tau_j^z + \gamma_j[1+n(E_j)]\left[\tau_j^+\rho(t)\tau_j^-
    -\half\left\{\tau_j^-\tau_j^+,\rho_S(t)\right\}  \right] 
  + \gamma_j n(E_j)\left[\tau_j^-\rho(t)\tau_j^+
    -\half\left\{\tau_j^+\tau_j^-,\rho_S(t)\right\}
  \right]
\Bigg\}
\end{equation}
%
\end{widetext}
where
\begin{math}
\gamma_j \equiv 2\pi J_j(E_j)
\end{math} is the dephasing rate,
\begin{equation}
\label{entangle::eq:10}
\delta E_j \equiv \Pr\int_0^{\infty} dE\;
\coth\left(\frac{E}{2T}\right)\,\frac{J_j(E)}{E_j -E}
\end{equation}
is the environment-induced level shift \cite[]{Zwanzig60a}, 
and
\begin{math}
n(E) \equiv [e^{E/T}-1]^{-1}
\end{math}
is the Bose distribution function.  It is again straightforward to find
the solution $\rho_S(t)$ of Eqs. (\ref{entangle::eq:4}) and
(\ref{entangle::eq:9}), which gives the concurrence (for $\gamma_1 =
\gamma_2 \equiv \gamma$ and $T=0$)
\begin{equation}
\label{entangle::eq:15}
C(r, t) = \Theta(r-1/3)
e^{-2\gamma t}\left[ e^{\gamma t}r - e^{\gamma t} + \frac{1+r}{2} \right].
\end{equation}
It is observed that the concurrence again decays exponentially and the
rate is determined by the dephasing time of a separate spin.  For the
maximally entangled initial state, it is simply given by
\begin{equation}
\label{entangle::eq:14}
C(r{=}1, t) = e^{-2\gamma t} .
\end{equation}

%
We now go beyond the Markovian approximation (still within the first
Born approximation).  In the non-Markovian case, it is convenient to
take the Laplace transform
\begin{equation}
\tilde\rho_S(z) = \int_0^{\infty}dt\; e^{-zt} \rho_S(t)
\end{equation}
and solve the resulting algebraic equations in the place of
Eqs.~(\ref{entangle::eq:4}) and (\ref{master}).
Then the dissipative dynamics of the system is determined by the pole
structure of $\tilde\rho_S(z)$ on the complex $z$-plane
\cite{DiVincenzo04z}: Whereas an isolated simple pole at $z=z_k$ with
residue $r_k/2\pi{i}$ contributes an exponential part
\begin{math}
r_k e^{z_kt_k}
\end{math}
to the evolution of $\rho_S(z)$, a branch cut of $\tilde\rho(z)$
produces non-exponential behavior.

The dynamics of $\rho_S(t)$ in the non-Markovian limit is governed by
the correlation function
\begin{equation}
\label{bcor}
C_j(t)
= \int_0^{\infty} d\omega\; J_j(\omega)
\left[\coth\left(\frac{\omega}{2T}\right) \cos(\omega t) + i\sin(\omega t)\right]
\end{equation}
of the quantum Langevin force
\begin{math}
\sum_m [g_{j, m}b_{j, m}^\dag + g_{j, m}^* b_{j, m}].
\end{math}
In particular, the solution $\tilde\rho_S(z)$ involves the respective
Laplace transforms $\tilde{C}_j'(z)$ and $\tilde{C}_j''(z)$ of the real
and imaginary parts of $C_j(t)$, which are given explicitly by
\begin{equation}
\label{entangle::eq:11}
\begin{bmatrix}
\tilde{C}_j'(z)\\
\tilde{C}_j''(z)
\end{bmatrix} = -\frac{\alpha_jz}{2}
\begin{bmatrix}
\cos(z/\omega_c) & +\sin(z/\omega_c) \\
\sin(z/\omega_c) & -\cos(z/\omega_c)
\end{bmatrix}
\begin{bmatrix}
\textrm{Ci}(z/\omega_c)\\
\textrm{Si}(z/\omega_c)
\end{bmatrix}.
\end{equation}
Here $\mathrm{Si}$ and $\mathrm{Ci}$ are the sine- and the
cosine-integral, respectively.

\begin{figure}
\centering
\includegraphics[height=0.22\textheight]{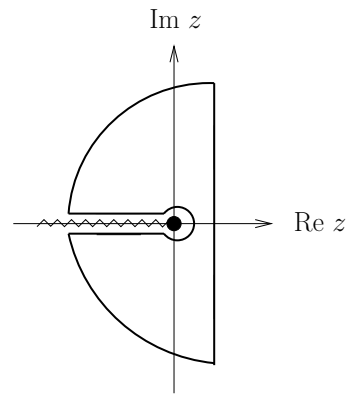}
\caption{Contour for the dephasing model.  The branch cut has been chosen 
for computational convenience.}
\label{depcon}
\end{figure}

For the pure dephasing model, the solution $\tilde\rho(z)$ involves only
$\tilde{C}_j'(z)$.  Note that the cosine integral $\textrm{Ci}(z)$ in
Eq.~(\ref{entangle::eq:11}) behaves as $\ln z$ in the limit $z\to 0$
while the sine integral is analytic.  This leads to a branch cut with no
isolated simple poles on the complex $z$ plane, as shown in
Fig.~\ref{depcon}.  It is remarkable that the solution has extra
structure not present in the Markovian approximation.  In particular,
the absence of isolated poles at $T=0$ indicates that there is no
exponential decay in $\rho_S(t)$ and the branch cut contribution gives
the leading correction.

\begin{figure}
\centering
\includegraphics[height=0.22\textheight]{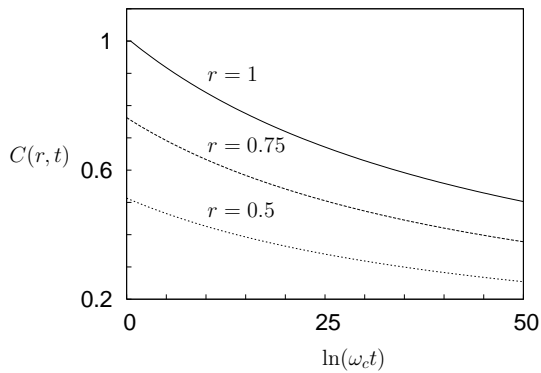}
\caption{Time evolution of the concurrence $C(r,t)$ for the pure
  dephasing model.}
\label{fig32}
\end{figure}

Figure~\ref{fig32} shows the time evolution of the concurrence 
for several different initial states ($r=1/2$, $3/4$, and $1$).
The intermediate and long time behaviors are given by $\ln \omega_c t$
and $(\ln \omega_ct)^{-1}$, respectively.
Specifically, for the maximally entangled initial state, we have
\begin{widetext}
\begin{equation}
C(r,t) = \frac{1}{8\pi\alpha}\left\{
  \left|(4\alpha+1)\pi
      -2\tan^{-1}\left[\frac{1-2\alpha\left(\gamma-\ln(\omega_c
            t)\right)}{2\pi\alpha}\right]\right|
- \left|(4\alpha-1)\pi
      +2\tan^{-1}\left[\frac{1-2\alpha
          \left(\gamma-\ln(\omega_c
            t)\right)}{2\pi\alpha}\right]
    \right|
\right\}.
\end{equation}
\end{widetext}
Note that unlike the Markovian case, the concurrence decays, albeit
slowly, even at zero temperature.

\begin{figure}
\centering
\includegraphics[height=0.22\textheight]{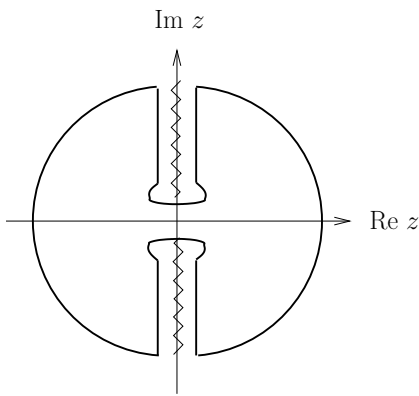}
\caption{Contour for the relaxation model.}
\label{relcon}
\end{figure}

For the relaxation model, the solution $\tilde\rho_S(z)$ involves
$\tilde{C}_j''(z)$ as well as $\tilde{C}_j'(z)$.  This leads to two
branch cuts on the complex $z$ plane, as shown in Fig.~\ref{relcon}.
Another difference from the pure dephasing model is the existence of an
isolated pole.  A closer investigation reveals that in the long-time
limit contributions from the branch cuts are negligibly small compared
with those from the isolated pole \cite{DiVincenzo04z}.  Accordingly,
$\rho_S(t)$ in the time domain decays exponentially, and the behavior of
the concurrence is similar to that in the Markovian limit.

%
In conclusion, we have used the generalized master equation to study
quantum dissipative dynamics of entanglement in the spin-boson model.
The pure dephasing model and the relaxation one have been considered,
and entanglement for both models has been shown to decay exponentially
in the Markovian limit.  When non-Markovian contributions are taken into
account, on the other hand, logarithmic decay has been revealed for the
dephasing model.

%
This work was supported by the BK21 Program, 
the SRC/ERC program (grant R11-2000-071),
the KRF Grant (KRF-2005-070-C00055), the SK Fund,
and the NCCR of Nanoscience in Switzerland.


\end{document}